\documentclass[aps,twocolumn,floatfix,superscriptaddress]{revtex4}
\usepackage{epsfig}

\begin{document}

\title{Growth of a vortex polycrystal in type II superconductors}

\author{Paolo Moretti}
\affiliation{Dipartimento di Fisica,
Universit\`a "La Sapienza", P.le A. Moro 2, 00185 Roma, Italy}
\affiliation{Center for Materials Science and Engineering,
University of Edinburgh,
King's Buildings, Sanderson Building, Edinburgh EH93JL, UK}
\author{M.-Carmen Miguel}
\affiliation{Departament de F\'{\i}sica Fonamental,
Facultat de F\'{\i}sica, Universitat de Barcelona, Av. Diagonal 647,
E-08028, Barcelona, Spain}
\author{Michael Zaiser}
\affiliation{Center for Materials Science and Engineering,
University of Edinburgh,
King's Buildings, Sanderson Building, Edinburgh EH93JL, UK}
\author{Stefano Zapperi}
\affiliation{INFM UdR Roma 1 and SMC, Dipartimento di Fisica,
Universit\`a "La Sapienza", P.le A. Moro 2, 00185 Roma, Italy}

\begin{abstract}
We discuss the formation of a vortex polycrystal in type II
superconductors from the competition between pinning and elastic
forces. We compute the elastic energy of a deformed grain boundary,
that is strongly non-local, and obtain the depinning stress for weak
and strong pinning. Our estimates for the grain size dependence on the
magnetic field strength are in good agreement with previous
experiments on NbMo. Finally, we discuss the effect of thermal noise
on grain growth.
\end{abstract}
\maketitle

Understanding the phase diagram of high temperature superconductors is
still a formidable challenge of modern condensed matter physics.
Typically high $T_c$ materials behave in a magnetic field as type II
superconductors, with further complications due to the broader phase
space --- in terms of temperature $T$ and field $H$ --- in comparison
to conventional superconductors \cite{BLA-94,BRA-95,GIA-01}.  Raising
the temperature, the Abrikosov vortex lattice \cite{ABR-57} melts into
a liquid \cite{SAF-92,BOC-01,AVR-01}, while quenched disorder leads to
more complex phases such as the vortex glass \cite{FIS-91} or the Bose
glass \cite{NEL-00}.  The natural question posed to the theorist is to
explain the occurence of the various vortex phases, linking the
experimentally observed behavior with the material microstructure.

The Bitter decoration technique provides a powerful method to
investigate the geometrical and topological properties of vortex
matter by direct imaging of the vortices \cite{GRI-94}. Its
application to conventional superconductors provided the first direct
evidence of the vortex lattice \cite{ESS-67} predicted by Abrikosov
\cite{ABR-57}. The lattice structure is often observed to coexist with
topological defects, such as isolated dislocations, dislocation
dipoles and grain boundaries. These last extended defects are the
signature of a vortex polycrystal with crystalline grains of different
orientations \cite{GRI-94,GRI-89}.  Vortex polycrystals have been
observed, after field cooling, in various superconducting materials
such as NbMo \cite{GRI-94,GRI-89}, NbS$_2$
\cite{MAR-97,MAR-98,FAS-02}, BSSCO \cite{LIU-94} and YBCO
\cite{HER-00}. The grain size is typically found to grow with applied
magnetic field \cite{MAR-97,GRI-89}.  Moreover, two-sided decoration
experiments show that the grain boundaries thread the sample from top
to bottom \cite{MAR-97,MAR-98}, i.e., one observes a columnar grain
structure.  Despite the wealth of experimental observations, there is
no detailed theory accounting for the formation of vortex
polycrystals. The issue is particularly interesting since recent
experiments indicate that the reentrant disordered vortex phase of
NbS$_2$, commonly believed to be amorphous, is instead polycrystalline
\cite{FAS-02}. Whether this is a stable thermodynamic phase is still
an open question.

The theoretical description of vortex matter in high $T_c$
superconductors is centered on the role of quenched disorder.
Early theoretical considerations seemed to imply that even a small
amount of disorder would lead to the loss of long-range order
\cite{LAR-70} and to the formation of an amorphous vortex glass
phase \cite{FIS-91}. Experimental observations did not confirm
this view, since ordered vortex structures are typically observed
even in presence of disorder. This contradiction was resolved by a
more detailed theoretical analysis of the weak disorder limit,
showing a topologically ordered, long-range correlated phase,
termed the Bragg glass \cite{GIA-95}. While the presence of a
Bragg glass has been confirmed experimentally \cite{KLE-01}, the
precise  nature of the transitions into the amorphous and liquid
phases is still debated. Recent theories highlight the importance
of dislocations as mediators of the transition \cite{KIE-00}, in
contrast with more traditional melting theories, based on the
Lindeman criterium \cite{BLA-94}. The properties of dislocations
in the vortex lattice have been the object of extensive
theoretical investigations \cite{NAB-80,BRA-86,MIG-97,KIE-00b}, but less
is known about grain boundaries.

\begin{figure}
\centerline{\psfig{file=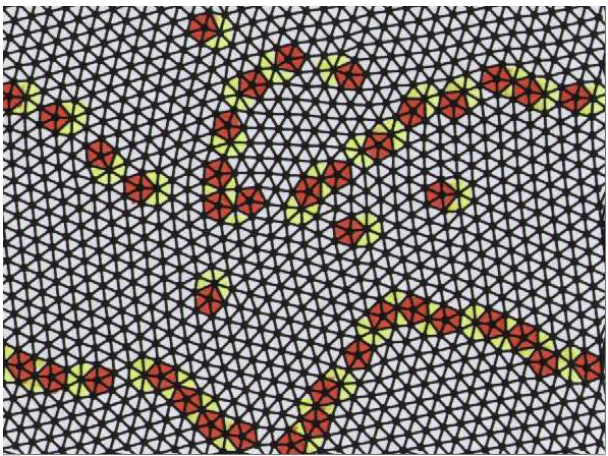,width=8cm,clip=!}}
\caption{Grain boundary structure from a simulation of inteacting
vortices after a sudden field cooling from a disordered vortex state.
The colored non six-fold coordinated vortices indicate dislocations, mostly
arranged into grain boundaries.}
\label{fig:0}
\end{figure}

Here we address the problem of the formation of a vortex polycrystal
from the point of view of grain growth. In field cooling experiments,
magnetic flux is already present in the sample as it is quenched in
the mixed phase. Thus it is reasonable to expect that vortices are
originally disordered and that, due to their mutual interactions,
undergo a local ordering process through the growth of grains with
various orientations which, in turn, implies the annihilation of
several dislocation lines as well as their organization into grain
boundaries separating the crystalline grains (see
Fig.~\ref{fig:0} for an example). The effect of quenched disorder is to pin the grain
boundaries, hindering the growth process. Thus to understand the
properties of vortex polycrystals, we analyze the dynamics of grain
boundaries in vortex matter as they interact with disorder.

We first use continuum anisotropic elasticity to evaluate the elastic
response of an extended three dimensional grain boundary to small
perturbations. In the large-wavelength limit, the grain boundary
elastic energy is found to be strongly non-local, with a linear wave
vector dependence in Fourier space. Hence, it is not feasible to use a
surface-energy approximation similar to the line-tension approximation
frequently used to describe the elastic response of isolated
dislocations. We thus use the non-local elastic energy in the
framework of weak and strong pinning theories and estimate the grain
size of a vortex polycrystal. Our results are in good agreement with
the experiments, reproducing the dependence of the grain size on the
applied magnetic field found in NbMo single crystals. In addition, we
discuss the effect of thermal activation and the associated creep
laws.

The elastic energy of the vortex lattice can be expressed in terms
of the vortex displacement field ${\bf u}$
\begin{equation}
\mathcal{H} =
\frac{1}{2}\int d^3r[c_{11}({\bf\nabla}{\bf u})^2+(c_{11}-c_{66})({\bf\nabla}
\cdot{\bf u})^2+c_{44}(\partial_z{\bf u})^2],\label{eq:elast}
\end{equation}
where $c_{11}$, $c_{44}$ and $c_{66}$ are respectively the
compression, tilt and shear moduli, and we assume the applied field to
point along the $z$ direction.  In the simplest continuum and
non-dispersive approximation the elastic moduli can be estimated as
$c_{11}\simeq c_{44} \simeq B^2/4\pi$ and $c_{66}\simeq \Phi_0
B/(8\pi\lambda)^2$, where $B$ is the magnetic induction, $\Phi_0$ is
the magnetic flux quantum and $\lambda$ is the London penetration
length.

We describe an extended grain boundary as an infinite array of edge
dislocations arranged regularly along the $y$ axis, with spacing $D$
and Burgers vector in the $x$ direction. Note that due to the columnar
grain structure of the vortex polycrystal, tilt and mixed boundaries
which would require a more complicated description in terms of screw
and edge dislocations, cannot occur. To obtain the elastic response of
the grain boundary to small perturbations, we consider a generic
deformation $\bf{v}= v_n(z) \hat{x}$ for the $n$th dislocation along
its glide plane and find the elastic displacements solving the elastic
equation associated with Eq.~(\ref{eq:elast}) with the appropriate
constraints induced by the dislocations. The grain boundary elastic
energy can then be obtained as in Ref.~\cite{MIG-97} by a suitable
expansion of Eq.~(\ref{eq:elast}). The full calculation is somewhat
involved and will be reported in detail elsewhere. Here we just quote
the resulting elastic energy which in Fourier space reads
\begin{equation}
\mathcal{H}_{GB}=\frac{\pi b^2}{2D^2}\sum_{G_y}\int\frac{dQ_y}{2\pi}
\int\frac{dk_z}{2\pi}\, M(Q_y+G_y, k_z)|\tilde{v}|^2, \label{eq:elast_gb}
\end{equation}
where the sum is over the reciprocal vectors $G_y\equiv 2\pi n/D$,
$|\tilde{v}|^2=\tilde{v}(Q_y,k_z)\tilde{v}(-Q_y,-k_z)$ and the
interaction kernel is given by
\begin{eqnarray}
M(k_y,k_z)&=&2c_{66}|k_y|+\frac{c_{66}^2}{c_{44} k_z^2}
\left[\frac{\left(2k_y^2+\frac{c_{44}}{c_{66}}k_z^2\right)^2}
{\sqrt{k_y^2+\frac{c_{44}}{c_{66}}k_z^2}}\right.\\
&-&\left. 4k_y^2\sqrt{k_y^2+\frac{c_{44}}{c_{11}}k_z^2}-2\left(\frac{c_{44}}{c_{66}}
-\frac{c_{44}}{c_{11}}\right)
|k_y|\,k_z^2\right].\nonumber
\end{eqnarray}
The long-distance behavior of the kernel is captured by the
behavior at small wave vectors that is given by
\begin{equation}
M(k_y,k_z)\simeq 2c_{66}|k_y|+\sqrt{c_{44}c_{66}}|k_z|,
\end{equation}
corresponding to a long-range non-local interaction in real space.  It
is convenient to work in an istropic reference frame, rescaling the
$y$ coordinate by a factor $\frac{1}{2}\sqrt{\frac{c_{44}}{c_{66}}}$
\cite{BLA-94}.  The elastic kernel then becomes $M({\bf k})\simeq
K|{\bf k}|$, with $K\equiv\sqrt{c_{44}c_{66}}$.  In thin films, we can
neglect the deformations along $z$ and the kernel is simply given by
$M(k)\simeq 2c_{66}|k|$. Note that for an isolated vortex lattice
dislocation it is correct up to a logarithmic factor to approximate
the elastic energy by an effective line tension \cite{MIG-97}, but a
similar procedure is not possible for a grain boundary.

Quenched disorder induces elastic deformations of the grain
boundaries and the competition between elasticity, disorder and a
driving force acting on the boundary can be analyzed in the
framework of pinning theories \cite{BLA-94,LAR-70}. Driving forces
for grain boundary motion can be externally induced by a current
flowing in the superconductor \cite{KIE-00b} or, as in our case,
internally generated by the ordering process during grain growth
\cite{HAZ-90}. Grain growth is driven by a reduction in energy:
For an average grain size $R$ and straight grain boundaries, the
characteristic energy stored per unit volume in the form of grain boundary
dislocations is of the order of $\Gamma_0/R$, where $\Gamma_0$ is the
energy per unit area of a grain boundary. Hence, the
energy gain achieved by increasing the grain size by $dR$ is 
$\Gamma_0/R^2dR$. 

Physically, the removal of grain boundary dislocations occurs
through the motion of junction points in the grain boundary
network. As junction points must drag the connecting boundary with
them, which may be pinned by disorder, motion can only occur if
the energy gain at least matches the dissipative work which has to
be done against the pinning forces. The dissipative work per unit
volume expended in moving all grain boundaries by $dR$ is
$\sigma_c b/(DR)dR$, where $\sigma_c$ is the pinning force per
unit area. Balancing against the energy gain yields the limit
grain size
\begin{equation}
R_g \approx \frac{D \Gamma_0}{b \sigma_c}\;.\label{eq:rg}
\end{equation}
In order to obtain an explicit expression for the grain size we
distinguish between weak and strong pinning regimes.

{\it Weak pinning:} In this regime, the grain boundary interacts
with random stresses $\sigma_{xy}$ induced by the elastic
deformations of the vortex lattice \cite{KIE-00,KIE-00b}. The
random stress correlations can be obtained directly from the
vortex displacement correlations in the random manifold (RM) and
Bragg glass (BG) regimes \cite{KIE-00}
\begin{eqnarray}
\overline{\sigma_{xy}({\bf r})\sigma_{xy}(0)}\approx
K^2\,\frac{a^2}{r^2}\left\{
\begin{array}{cc}
(r/R_a)^{2\zeta_{RM}} & \mbox{RM}\\
1 & \mbox{BG}
\end{array}
\right.\end{eqnarray} where $R_a$ is the length at which the
vortex displacements become of the order of the lattice spacing
$a$ and corresponds to the onset of the BG regime. The roughness
exponent in the RM is estimated as $\zeta_{RM} \approx 1/5$. The
pinning energy for the grain boundary is then given by
$\mathcal{H}_{pin}=\sum_n\int dz \,v(nD,z)\,b\sigma_{xy}[v,nD,z]$.
The roughness of the grain boundary can be obtained by a scaling
argument \cite{KIE-00b} comparing this expression with the elastic
energy in Eq.~(\ref{eq:elast_gb}), yielding
$\zeta_{GB}=\zeta_{RM}$ in the RM regime and a logarithmic
roughness for $R>R_a$. Notice that both isolated dislocations and
``dislocation bundles'' are found to be rougher than grain
boundaries \cite{KIE-00b}, which implies that the latter are more
stable.

\begin{figure}
\centerline{\psfig{file=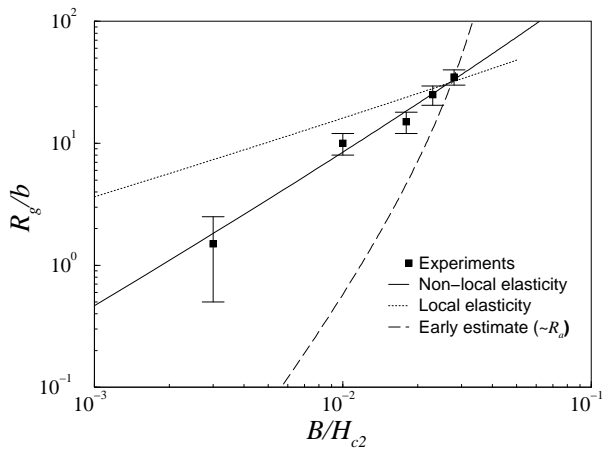,width=8cm,clip=!}} 
\caption{The grain size of a vortex polycrystal experimentally
obtained from Bitter decoration of a NbMo sample as a function of the
applied magnetic field (Ref.~\protect\cite{GRI-89}) is compared with
the theoretical predictions. The calculation based on Friedel
statistics with non-local elasticity compares favorably with the
data. For comparison we report as well the result obtained using
Friedel theory with local elasticity and the earlier estimate from
Ref.~\protect\cite{GRI-89}, formally equivalent to our weak pinning
result.} \label{fig:1}
\end{figure}

The depinning stress can be computed within the framework of
collective pinning theory: the energy associated with bending a grain
boundary fraction of linear dimension $L < R_a$ over the
characteristic distance $v$ can be estimated as
\begin{equation}
\mathcal{E}=\frac{Kb^2}{
D^2}Lv^2-\frac{Kab}{D}Lv\left(\frac{L}{R_a}\right)^{1/5}+\sigma b L^2v/D,
\end{equation}
where the first term represents the elastic energy, the second the
pinning energy, and the third the work done by an external driving
stress $\sigma$ in displacing the boundary. Minimizing the first
two terms, for $v=a \simeq b$ \cite{KIE-00b}, 
we obtain the ``plastic'' Larkin length $L_p
\simeq (b/D)^5 R_a$, which is typically smaller than $R_a$.
The depinning stress is identified as the stress necessary to
depin a section of dimension $L_p$: $\sigma_c=Kb^2/(D L_p)$.
Combining this expression with Eq.~(\ref{eq:rg}), using
$\Gamma_0 \simeq Kb^2/D$, we obtain $R_g \sim R_a$. The identification of
$R_g$ with $R_a$ was proposed in Ref.~\cite{GRI-89}, but was not
confirmed by experiments (see Fig.~\ref{fig:1} and
Ref.~\cite{GRI-89}) \cite{nota-GRI}. We therefore propose to
interpret the experimental data under a strong pinning assumption.

{\it Strong pinning:} In this regime, pinning centers are strong and
localized and one can assume that the dislocations forming the grain
boundary are pinned by individual obstacles. We consider here the case
of columnar defects, oriented along the $z$ axis. This case should be
relevant for the experiments of Ref.~\cite{GRI-89} where grain
boundary pinning is provided by screw dislocations in the
superconducting crystal. In these conditions, the problem becomes
effectively two dimensional (2D) and we can directly generalize the
strong pinning theory of Friedel \cite{Friedel}, developed in the
context of single dislocations, by taking into account non-local
elastic interactions. The basic idea is to consider a grain boundary
segment as it depins from a pair of strong obstacles. The length $L$
of the segment corresponds to the effective spacing between obstacles
along the grain boundary, and it forms a bulge of maximum width
$v$. After the grain boundary segment overcomes the pin it will travel
by an amount which is, again, of the order of $v$ and, hence, sweep an
area of the order of $L v$.  At the depinning threshold, the grain
boundary starts to move through a sequence of statistically equivalent
configurations, and the freed segment will encounter, on average,
precisely one new obstacle in the course of this process. This
argument leads to the condition $Lv \simeq 1/\rho$ where $\rho$ is the
area density of pinning defects. The elastic energy per unit length of
the bulge of width $v$ and extension $L$ is $2c_{66}b^2 v^2/D^2$ (2D
result), and should balance the work per unit length $\sigma bLv/D$
done by the driving stress $\sigma$ in bowing the boundary. This
energy balance provides a relation between $L$ and $v$. Furthermore,
at depinning the total force $d\sigma b L/D$ should be equal to the
defect strength $f_0$, where $d$ is the sample thickness. Combining
the equations above we obtain the depinning stress $\sigma_c b=D
f_0/(d L_f)$, where the Friedel length $L_f$ is given by $L_f=2c_{66}
b^2 d/(f_0\rho D^2)$. Inserting the expression for the critical stress
in Eq.~(\ref{eq:rg}) together with the scale-independent surface
tension $\Gamma_0= 2c_{66}b^2/D$, we obtain
\begin{equation}
\frac{R_g}{b} \approx \frac{c_{66}^2 b^3 d^2}{D^3f_0^2\rho}.
\end{equation}
In order to use this result to fit the data in Ref.~\cite{GRI-89}, we
have to express it in terms of the reduced field $\tilde{B}\equiv
B/H_{c2}$, where $H_{c_2}$ is the upper critical field of the
superconductor. The field dependence is implicit in the parameters $b$
and $D$, i.e. $b \sim D \sim a \sim \tilde{B}^{-1/2}$, as well as in
the shear modulus $c_{66} \sim \tilde{B}$, and in the pinning strength
$f_0$. The pinning force due to a screw dislocation was computed in
Ref.~\cite{SCH-78} and is given by
$f_0\propto\tilde{B}^{1/2}(1-\tilde{B})\ln(\xi/2.7b_0\tilde{B})
\approx \tilde{B}^{1/2}\ln(\xi/2.7b_0\tilde{B})$, where
$\xi \simeq 100\AA$ is the coherence length \cite{KRU-91}, and
$b_0\simeq 5\AA$ is the Burgers vector of the screw dislocation
\cite{SCH-78}. The resulting expression predicts a linear field
dependence of the grain size with logarithmic corrections. In
Fig.~\ref{fig:1} we can corroborate that the agreement of this
prediction with magnetic decoration data from Ref.~\cite{GRI-89} is
quite satisfactory, especially if compared to the estimate based on
local elasticity assumptions.

In the discussion above we have neglected thermal fluctuations, which
could induce an activated motion of the grain boundaries, particularly
in high $T_c$ materials. This problem can be approached generalizing
scaling theories of creep for vortices and dislocations
\cite{BLA-94,KIE-00b}. In the weak pinning regime, the relevant energy
barrier that the grain boundaries have to surmount under an applied
stress $\sigma <\sigma_c$ is given by $U(\sigma) =
U_0(\sigma_c/\sigma)^\mu$, where $U_0\simeq K b^3 R_a$ and $\mu=1$. In
our case, the applied stress is the ordering stress, so that we have
$\sigma_c/\sigma \simeq R/R_a$. Using this expression in the energy
barrier for thermally activated grain growth, it follows
\begin{equation}
\tau\frac{dR}{dt}=R_a \exp{\left[-\frac{U_0}{kT}\frac{R}{R_a}\right]},
\end{equation}
where $\tau$ is the appropriate characteristic time. The equation
can readily be solved yielding, in the long time limit, a
logarithmic growth $R(t)/R_a=kT/U_0 \log(t/\tau)$. This law holds
for $R>R_a$ when the grain boundaries would be pinned at $T=0$. In
the initial growth stage $R\ll R_a$, we can neglect pinning forces
and the dynamics is ruled by the ordering stress: $\dot{R} \sim
1/R$, yielding a power law growth $R(t) \sim \sqrt{t}$.

In conclusion, we have analyzed the problem of grain growth in a
vortex polycrystal studying the dynamics of grain boundaries. We have
obtained estimates for the grain size that compare well with
experiments on NbMo crystals, and derived the law for thermally
activated grain growth. In general, our theory should apply to field
cooling experiments in which there is a competition between the
ordering stress and the pinning stress. The resulting polycrystalline
structure, at least in the weak pinning regime, represents a
metastable state in which the system is trapped during its evolution
towards the stable Bragg glass phase. The possibility of a
thermodynamically stable polycrystalline vortex state, recently
suggested by experiments \cite{FAS-02,MEN-03}, still remains to be
confirmed theoretically.


\begin{thebibliography}{99}
\bibitem{nota-GRI}
We have reached this result from  a completely different
perspective than Ref.~\protect\cite{GRI-89}, where
it is assumed that above $R_a$ long-range order is lost in the
equilibrium state, while it is now known that this is not the case
\protect\cite{GIA-95}. Our result describes instead
a metastable state due to the arrested grain growth at $T=0$.
\end{thebibliography}
\end{document}